\documentstyle[12pt]{article}

\begin{document}

\begin{titlepage}

\begin{center}
\baselineskip 24pt
{\Large {\bf Possible Explanation for $E > 10^{20}$ eV Air Showers with
Flavour-Changing Neutral Currents}}\\
\vspace{.5cm}
\baselineskip 16pt
{\large Jos\'e BORDES}\\
\vspace{.2cm}
{\it Dept. Fisica Teorica, Univ. de Valencia,\\
  c. Dr. Moliner 50, E-46100 Burjassot (Valencia), Spain}\\
\vspace{.3cm}
{\large CHAN Hong-Mo, Jacqueline FARIDANI}\\
\vspace{.2cm}
{\it Rutherford Appleton Laboratory,\\
  Chilton, Didcot, Oxon, OX11 0QX, United Kingdom}\\
\vspace{.3cm}
{\large Jakov PFAUDLER}\\
\vspace{.2cm}
{\it Dept. of Physics, Theoretical Physics, University of Oxford,\\
  1 Keble Road, Oxford, OX1 3NP, United Kingdom}\\
\vspace{.3cm}
{\large TSOU Sheung Tsun}\\
\vspace{.2cm}
{\it Mathematical Institute, University of Oxford,\\
  24-29 St. Giles', Oxford, OX1 3LB, United Kingdom}\\
\end{center}

\vspace{1cm}
\begin{abstract}
Air showers observed with $E > 10^{20}$ eV pose a problem since protons
at such energies cannot survive a long journey through the 2.7 K microwave
background.  It is suggested that they are initiated instead by neutrinos
which become strongly interacting through exchanges of gauge bosons
associated with generation-changing neutral currents, a suggestion which is 
supported by calculations in a dualized standard model recently proposed.
In turn, the suggestion bounds the mass of the gauge bosons and gives 
estimates for the branching ratios of FCNC s, c, b, and t meson decays 
which are accessible to experiments now being planned.
\end{abstract}

\end{titlepage} 

\clearpage

Over the last thirty years, evidence has accumulated for the existence of
air showers with primary energies greater than $10^{20}$ eV \cite{Volcano,
Haverah,Yakutsk,Flyseye,Agasa}.  To-date, 9 such events have been recorded 
by several detectors using a variety of detection techniques, which makes 
it rather unlikely for all of them to be due to experimental biases or
errors.  Their existence poses an intriguing physics question and it is 
mainly to investigate further these so-called EHECR's (extremely high 
energy cosmic rays) that the huge Auger project is being planned, involving 
large arrays on two sites, one in each hemisphere, totalling in area 
6000 km$^2$ \cite{Auger}.  The reason for this unusual amount of interest 
is that air showers with such energies ought not in theory to be there at 
all.  Air showers at high energies are thought to be initiated mostly by 
protons, and protons at such an energy would quickly lose it by interacting 
with the photons in the 2.7 K microwave background field via, for example, 
the interaction:
\begin{equation}
p + \gamma_{2.7 K} = \Delta + \pi.
\label{pgammaint}
\end{equation}
Indeed, it has been shown by Greisen \cite{Greisen} and by Zatsepin and 
Kuz'min \cite{Zatmin} that the spectrum of protons originating from more 
than 50 Mpc away should be cut off sharply at around $4 \times 10^{19}$ eV 
in traversing the microwave background field. Thus, the observed air showers 
with energies in excess of $10^{20}$ eV at experimental energy resolutions 
of order 20 percent would be a blatant contradiction to theoretical 
expectations unless their primary protons have an origin within that sort
of distances.  However, such nearby sources are thought to be unlikely 
for the following reason.  Over such short distances, protons with these
extreme energies will be hardly deflected by the magnetic fields either
in our galaxy or in the space between.  The observed directions of the
air showers should thus point directly to their sources, but no candidate
sources have been found within a distance of 50 Mpc which are thought 
capable of producing particles of such an enormous energy.

A possible alternative explanation for these showers can be that they are 
not initiated by protons at all but by some other particles.  Thus a stable 
zero-charged particle, such as a neutrino, could survive the long journey 
from whatever its extragalactic origin through the microwave background 
to arrive on earth with its high energy intact \cite{Sigljee,Halzen,Elmers}.
However, a neutrino with only the known electroweak interactions 
can readily penetrate our atmosphere.  In order to interact with the
air at all to produce air showers at the observed frequency, cosmic 
neutrinos at these energies would need to have a very large flux, which 
is hardly imaginable.  Even if this high flux is indeed available, the
air showers induced would have angular and depth distributions which are
at variance with those observed.  Whereas neutrino initiated showers are
expected to be mostly horizontal, in order that the neutrino may pass 
through sufficient air for it to effect a collision, the observed events 
(with angular resolution of only $1^o - 2^o$), nearly all have incident 
angles of less than $40^o$ from the zenith.  Further, air showers initiated
by weakly interacting neutrinos would have a flat distribution in the depth
of atmosphere penetrated, not bunched at high altitudes as observed.  One 
concludes, therefore, that if neutrinos have only the known electroweak 
interactions, then the observed air showers with energies greater than 
$10^{20}$ eV are very unlikely to be initiated by neutrinos.

On the other hand, if neutrinos have interactions which become strong 
at ultra-high energies then the objections raised in the paragraph above
are overruled and neutrinos can afford an explanation for the events under 
consideration.  This conjecture has been considered in \cite{Sigljee,Halzen,
Elmers} on the basis of possible substructures to quarks and leptons yet 
unknown to us \cite{Domokov,Domokos}.  What we wish to point out in this 
note is that there is another theoretical scenario which will naturally 
lead to such interactions.  This goes as follows.

Neutrinos, like other leptons and quarks, are known to exist in three 
families or generations.  This fact has no explanation in the conventional
formulation of the standard model but is merely introduced into the theory 
as a phenomenological requirement.  Now, a favourite suggestion among 
theoreticians is that generations may in fact represent the quantum numbers 
of a broken continuous symmetry like $SU(3)$.  To bring it into line with 
other known continuous symmetries, we would then want this new one to be 
also a gauge symmetry.  If so it has to be mediated by a new set of gauge 
bosons and these, being flavoured but uncharged, would lead in turn to flavour
changing neutral currents (FCNC).  Now, such gauge bosons will have to be
very heavy, for otherwise they will give rise to sizeable FCNC decays, which 
have not been observed.  Indeed, the strongest bounds on the gauge boson mass 
coming from $K$-decays are usually given to be in the 10 - 100 TeV region,
depending on the strength of the gauge coupling \cite{Cahnrari}.  
If we accept this scenario, then at energies below, say, 100 TeV, 
generation-changing interactions due to the exchange of these 
bosons will be negligible and neutrinos will interact just via the usual 
electroweak forces.  However, at energies greater than 100 TeV, the new 
forces, which could in principle be strong, will come into play and give 
rise to new effects.

The incoming primary energy of the air showers under consideration is of
order $10^{20}$ eV, which in collision with a proton in the air corresponds
to a CM energy of around 400 TeV.  They are therefore, according to the
estimates of the preceding paragraph, at an energy possibly above the 
advent of the new interactions.  Hence, neutrinos at these energies may have 
already acquired strong interactions and can give rise to the observed air 
showers as required.

A particular realization of this theoretical scenario is afforded by a
recently proposed scheme based on a nonabelian generalization of 
electric-magnetic duality \cite{Chantsou}.  In this scheme, the generation 
index is identified with dual colour, from which assumption it follows 
that there are exactly three generations and that the generation 
symmetry is broken.  The symmetry is mediated by the dual gluons whose 
couplings are related to the usual couplings of colour gluons by the Dirac 
quantization condition, which in the standard conventions used in nonabelian
theories with $\alpha = g^2/4\pi$ reads as:
\begin{equation}
{\tilde g} g = 4 \pi,
\label{Diraccond}
\end{equation}
and are seen at these energies to be large.  Furthermore, the CKM matrix 
in this scheme is the identity matrix at tree level and only acquires 
mixing from loop corrections, which, being dependent on the symmetry 
breaking scale, allow one to make an estimate of the dual gluon mass through 
the observed magnitude of the mixing elements.  A preliminary result from 
a calculation along these lines, the details of which will be reported 
elsewhere, gives an estimate of the dual gluon mass of around several 100 
TeV.  Hence, in this scheme, not only is it possible for the neutrino to 
acquire strong interactions at energies above around 100 TeV as suggested 
in the general framework outlined in the above paragraph, but it seems that 
it is even {\it predicted} to be so.  If that is indeed the case, then air 
showers initiated by neutrinos with energies greater than $10^{20}$ eV 
would occur so long as neutrinos with such energies are produced somewhere 
out there in the universe.

Are there viable sources?  Since neutrinos are supposed to interact strongly
at such energies, then any source capable of accelerating protons to these
energies can produce neutrinos directly from collisions of the accelerated 
protons, a mechanism seemingly more efficient for high energy neutrinos 
than by, for example, pion decay.  Now, of the three possible candidate 
categories of sources lying above the line:
\begin{equation}
BR = E/Z,
\label{Hillas}
\end{equation}
on the Hillas plot \cite{Hillas} (where $B$ is the magnetic field in 
$\mu$G, $R$ the size in kpc, $E$ the energy in EeV = $10^{18}$ eV, and 
$Z = 1$ for protons), which are thought capable of accelerating protons to 
these energies, two are thought to have difficulty emitting them 
\cite{Boratav}.  For the neutron star, the accelerated proton is liable 
to lose its energy by sychrontron radiation on escaping simply by crossing 
the magnetic field which is itself responsible for its acceleration.  On 
the other hand, for active galactic nuclei, the accelerated proton is 
expected to suffer energy loss in its escape by interacting with the 
intense radiation field thought to surround the central parts of the AGN.  
We notice, however, that neither of these effects would affect the neutrino, 
which being neutral would not interact electromagnetically and would thus 
be able, once it is produced by the mechanism suggested above, to escape 
with its energy intact.

A neutrino interacting strongly at extreme energies would even offer
possible answers to several puzzling questions connected with the origin 
of $E > 10^{20}$ eV air showers.  For instance, three pairs among the observed
showers are known to have a common direction to within $2^o$ \cite{Hayashida}, 
suggesting thus a common origin for each pair.  However, if they are charged 
particles and have different energies as these pairs do, then they ought to be 
deflected differently by the intervening magnetic fields and arrive with
different directions unless the sources are rather close to earth.  This 
objection, however, does not apply to neutrinos so that each pair could
have come from the same distant source.  Further, it has been noted 
that the highest energy event known, namely the 320 EeV event recorded by
the Fly's Eye detector \cite{Flyseye}, points in the direction of a very
powerful Seyfert galaxy (MCG 8-11-11) which is 900 Mpc away \cite{Elmers}.  
If this is taken to be the source of that particular event, then one may 
wonder why such a source capable of producing a 320 EeV particle should 
give no signal in the 10 EeV range, which could be easily detected by
the Fly's Eye detector \cite{Boratav}.  This objection, however, poses 
no difficulty for the neutrino which interacts strongly only at extreme 
energies well above 100 TeV CM.  At lower energy, the interaction being 
there supposedly weak, neutrinos cannot, first of all, be produced directly 
from the collision of high energy protons as suggested above, and secondly, 
even if some of them are produced in MCG 8-11-11, the $\nu N$ cross 
section would have decreased sufficiently by these energies as to give 
them little chance of initiating air showers when they arrive on earth. 

Now, if such neutrinos are produced, by MCG 8-11-11 or some such object, then
they will be able to reach us.  They will be attenuated by neither the 2.7 K 
background photons since they are chargeless, nor by the 1.9 K background 
neutrinos, if massless, since their collisions will have CM energies of 
only around 200 MeV (even a 10 eV neutrino will give only 40 GeV CM energy)
at which the interaction is still very weak.  But would they be able to 
produce showers with the observed properties, such as the above-mentioned 
angular and depth distributions?  The answer to this question depends on 
neutrino cross sections at $10^{20}$ eV which one is at a loss how to 
estimate in the general framework of generation-changing neutral currents 
with unknown couplings.  However, in the scheme based on duality suggested 
in \cite{Chantsou}, one has a slight handle in that the coupling of the 
dual gluon is given by the Dirac quantization condition (\ref{Diraccond})
above.  At low energies, the neutrino cross section due to
dual gluon exchange will behave as a point cross section, similarly to that
from $W$-exchange, rising linearly with the beam energy.  If one takes
naively the experimental $\nu N$ cross section extrapolated to the $W$
mass and scales that by ${\tilde \alpha}_3^2/\alpha_2^2$, with $\alpha_2$
being the usual electroweak coupling and ${\tilde \alpha}_3$ the dual
colour coupling obtained from (\ref{Diraccond}), one obtains a cross 
section at the dual gluon mass of around 10 $\mu$b, approaching the 
hadronic range.  However, the dual colour coupling being very large, 
${\tilde \alpha}_3 \sim 16$, it is doubtful whether the above estimate 
has much significance.  

There is yet another mechanism which can give sizeable cross sections to 
extreme energy neutrinos for producing showers.  An antineutrino hitting
an electron in the atmosphere at $10^{20}$ eV primary energy gives a 
CM energy of around 10 TeV.  If the spectrum of the generation-changing
gauge and their associated Higgs bosons is populated at this mass range, 
then ${\bar \nu}$ and $e$ can combine to form one of these states which 
will then quickly decay into hadrons and initiate a shower.  This can 
happen so long as generation is a broken gauge symmetry, independently 
of the special dual scheme of \cite{Chantsou}.  However, with more knowledge 
obtained from the current calculations being done with the dual scheme, 
there may be a chance of specifying the boson spectrum and even of 
estimating this resonant cross section.

Provided one accepts as a working hypothesis that the observed air showers 
with greater than $10^{20}$ eV energies are initiated by neutrinos, then 
independently of the dual colour interpretation suggested in \cite{Chantsou}, 
one would obtain an upper limit of 400 TeV for the mass of the gauge bosons 
mediating generation-changing reactions.  Together with the lower limit 
of around 100 TeV obtained from the bounds on FCNC K-decays, these would 
limit the gauge boson mass within sufficiently narrow limits to make 
predictions of FCNC decays in other reactions meaningful.  In Table 1, we 
list the branching ratios so predicted, assuming a unique gauge boson
mass, for various FCNC decay modes of s, c, b and t particles which should 
be available for scrutiny at Daphne, Barbar and other strange-, charm-,
bottom- and top-factory experiments now being planned.  Calculations along
the lines of the dual scheme of \cite{Chantsou} will give more detailed 
predictions which we hope to report later.

\begin{table}

\begin{eqnarray*}
\begin{array}{||l|l|l||}
\hline \hline
  & Theoretical \, \, Estimate & Experimental \, \, Limit \\
Br(K^+ \rightarrow \pi^+ ll') & f_{s \rightarrow d l l'} 
\left(\frac{\tilde{g}^2}{4\pi}\right)^2 2 \times 10^{-12} &
2.1 \times 10^{-10}   \\
Br(K^0_s \rightarrow ll') & f_{sd \rightarrow l l'} 
\left(\frac{\tilde{g}^2}{4\pi}\right)^2 9 \times 10^{-11} &
3.2 \times 10^{-7}  \\
Br(D^+ \rightarrow \pi^+ ll') & f_{c \rightarrow u l l'} 
\left(\frac{\tilde{g}^2}{4\pi}\right)^2 2 \times 10^{-13} &
1.8 \times 10^{-5}  \\
Br(B^+ \rightarrow \pi^+ ll') & f_{b \rightarrow d l l'} 
\left(\frac{\tilde{g}^2}{4\pi}\right)^2 10^{-10} &
3.9 \times 10^{-3}  \\
Br(B^+ \rightarrow K^+ ll') & f_{b \rightarrow s l l'} 
\left(\frac{\tilde{g}^2}{4\pi}\right)^2 10^{-10} &
6 \times 10^{-5}  \\
\Gamma(t \rightarrow q ll') & f_{t \rightarrow q l l'} 
\left(\frac{\tilde{g}^2}{4\pi}\right)^2 9 \times 10^{9} s^{-1} &
 \\
\hline \hline
\end{array}
\end{eqnarray*}

\caption{The estimates given above assume a unique mass of 400 TeV for the
gauge bosons with the gauge coupling ${\tilde g}$.  The coefficients
$f$ involve the mixing angles but are bounded by and of order unity.
For the dual scheme of \protect\cite{Chantsou}, ${\tilde g}$ is given by the
Dirac quantization condition (\protect\ref{Diraccond}) in terms of the ordinary
colour gluon coupling run to 400 TeV, and corresponds to a value for
$({\tilde g}^2/4\pi)^2$ of around 250.  The resulting branching ratios
satisfy the present experimental bounds but are accessible to new experiments
now being planned.  Detailed calculations with nondegenerate gauge boson
masses and explicit $f$'s depending on mixing angles will be reported 
elsewhere.}

\end{table}

As so little is yet known about the envisaged new interactions for 
neutrinos at high energies, we cannot suggest at present decisive tests 
for confirming or rejecting, directly from measurement of air shower 
properties, the assumption that showers with $E > 10^{20}$ eV are indeed 
initiated by neutrinos.  However, a good test for the hypothesis already
exists in terms of the FCNC decays as discussed in the preceding paragraph 
and listed in Table 1.

\vspace{.2cm}

We thank Jeremy Lloyd-Evans for interesting us in high energy air
showers and for giving us some valuable information on the subject.


\begin{thebibliography}{99}

\bibitem{Volcano} J. Linsley, Phys. Rev. Lett. 10 (1963) 146.

\bibitem{Haverah} M.A. Lawrence, R.J.O. Reid and A.A. Watson, J. Phys. G,
   17 (1991) 773.

\bibitem{Yakutsk} B.N. Afanasiev et al., Proc. of the 24th ICRC, Rome,
   Italy, 2 (1995) 756.

\bibitem{Flyseye} D.J. Bird et al., Ap. J. 424 (1994) 491.

\bibitem{Agasa} S. Yoshida et al., Proc. of the 24th ICRC, Rome, Italy,
   1, (1995) 793.

\bibitem{Auger} The Pierre Auger Observatory Design Report (2nd ed.),
   14 March 1997.

\bibitem{Greisen} K. Greisen, Phys. Rev. Letters, 16 (1966) 748.

\bibitem{Zatmin} G.T. Zatsepin and V.A. Kuz'min, JETP Letters, 4 (1966) 78.

\bibitem{Sigljee} G. Sigl, D.N. Schramm and P. Bhattachajee, Astrop. Phys.
   2 (1994) 401.

\bibitem{Halzen} F. Halzen et al., Astrop. Phys. 3 (1995) 151.

\bibitem{Elmers} J.W. Elbert and P. Sommers, Ap. J. 441 (1995) 151.

\bibitem{Domokov} G. Domokos and S. Nussinov, Phys. Letters 187B (1987) 372.

\bibitem{Domokos} G. Domokos and S. Kovesi-Domokos, Phys. Rev. D38
   (1988) 2833.

\bibitem{Cahnrari} e.g. Robert N. Cahn and Haim Harari, Nucl. Phys. B176
   (1980) 135.

\bibitem{Chantsou} Chan Hong-Mo and Tsou Sheung Tsun, Rutherford Appleton
   Laboratory report, RAL-TR-97-005, hep-th/9701120.

\bibitem{Hillas} A.M. Hillas, Annual Review Astron. Astrosphys. 22 (1984)
   425.

\bibitem{Boratav} Murat Boratav, astro-ph/9605087.

\bibitem{Hayashida} N. Hayashida et al., Phys. Rev. Letters 77 (1996) 1000.

\end{thebibliography}
\end{document}